# Ultrafast transmission electron microscopy on dynamic process of a CDW transition in 1T-TaSe$_2$

Shuaishuai Sun[1*], Linlin Wei[1*], Zhongwen Li[1], Gaolong Cao[1], Y.Liu[2], W.J.Lu[2], Y. P. Sun [3, 4], Huanfang Tian[1], HuaixinYang[1†] and Jianqi Li[1, 5‡]

[1]*Beijing National Laboratory for Condensed Matter Physics, Institute of Physics, Chinese Academy of Sciences, Beijing 100190, China*

[2] *Key Laboratory of Materials Physics, Institute of Solid State Physics, Chinese Academy of Sciences, Hefei 230031, China*

[3] *High Magnetic Laboratory, Chinese Academy of Sciences, Hefei230031, China*

[4] *Collaborative Innovation Center of Advanced Microstructures, Nanjing University, Nanjing 210093, China*

[5]*Collaborative Innovation Center of Quantum Matter, Beijing 100084, China*

Four-dimensional ultrafast transmission electron microscopy (4D-UTEM) measurements reveal a rich variety of structural dynamic phenomena at a phase transition in the charge-density-wave (CDW) 1T-TaSe$_2$. Through the photoexcitation, remarkable changes on both the CDW intensity and orientation are clearly observed associated with the transformation from a commensurate (C) into an incommensurate (IC) phase in a time-scale of about 3 ps. Moreover, the transient states show up a notable "structurally isosbestic point" at a wave vector of q$_{iso}$ where the C and IC phases yield their diffracting efficiencies in an equally ratio. This fact demonstrates that the crystal planes parallel to q$_{iso}$ adopts visibly common structural features in these two CDW phases. The second-order characters observed in this nonequilibrium phase transition have been also analyzed based on the time-resolved structural data.

PACS numbers: 68.37.Lp, 61.05.jm, 64.70.Rh, 78.47.J-

One of the great experimental challenges in physics and material sciences is to obtain a real time view of the extremely fast changes in atomic configuration association with structural phase transitions and chemical reactions [1-2]. Recent developments on the ultrafast electron diffraction (UED) [3-4], ultrafast X-ray diffraction (UXRD) [5-6] and four-dimensional ultrafast transmission electron microscopy (4D-UTEM) [7-12] have been demonstrated as the effective techniques to reveal structural dynamics at the time resolution better than one picosecond. In addition to the extensive use of the TEM to characterize the local microstructure features, the capability obtained dynamic structural information from an area as small as a few nanometers to microns is an unparalleled strength of UTEM [9-12]. In this paper, we will report on a study of the CDW phase transition in 1T-TaSe$_2$ between a commensurate phase (C-phase, $q_C = 0.225a* + 0.07b*$) and an incommensurate phase (IC-phase, $q_{IC} = 0.278b*$) induced by the fs-laser excitation. Theoretical analysis of the topology of the Fermi surface suggests that CDW in the IC-phase can be connected by the modulation wave vector as predicted by Peierls model [13-14], however, the nature of the C-phase and CDW transitions, is still under debate [15-18]. It is believed that the CDW state in 1T-TaSe$_2$ often results in visible modulation of the electron density and associated with superstructure [13, 19]. In previous literatures, it is demonstrated that the layered 1T-TaSe$_2$ crystals contain numerous of remarkable structural and physical phenomena in correlation with CDW and superconductivity [20-22]. 1T-TaSe$_2$ materials have a quasi-two-dimensional structure consisting of planes of hexagonally arranged Ta atoms, sandwiched by two Se layers coordinating the central Ta atom in an octahedral arrangement [15]. The structural modulations in association with CDW states occur dominantly in the a-b plane which depends evidently on the temperature and chemical compositions [21]. We herein report on direct observations of the lattice dynamic properties through the CDW transition in 1T-TaSe$_2$ induced by the fs-laser excitation. Our UTEM measurements have revealed noticeably fast rotation of CDW vectors and, moreover, a notable "structurally isosbestic point" has been identified for the nonequilibrium CDW transient. Certain fundamental features for the structural transition and the threshold effects of photoexcitation for the CDW modulations have been also discussed in this layered system.

In order to achieve the necessary time and spatial resolution with complete structural information on dynamic properties of the transient processes in advanced materials, we have focused our attention on development of an UTEM at Institute of Physics, Chinese Academy of Sciences (IOP, CAS). Fig. 1a shows the photography of 4D-UTEM built at IOP. This UTEM has been developed based on a modification of a 200kV electron gun in which the electronic configuration and vacuum systems have been re-designed and improved. Our recent experimental measurements show this UTEM can be well performed under either the photoelectron emission mode for time-resolved observations or the thermionic mode as a conventional high-resolution TEM [12]. The laser system equipped with this UTEM is used to generate the electron pulses and to pump samples for stroboscopic observations; each experimental image in present study is constructed stroboscopically from typically $10^5$ pulses. The time durations between laser pulses have been tuned depending on samples to allow efficient heat dissipation in stroboscopic experiments. Single crystalline samples used in present study have been well characterized as reported in Ref. [21]. In our UTEM experiments, the 1T-TaSe$_2$ single crystals were mounted on well-aligned multi-walled carbon nanotubes (MWCNT) bunches which were textured and arranged in a decussating pattern on the TEM Cu-grids as shown in inserted image of Fig. 1a. It is demonstrated that this kind of MWCNT networks could result in a well thermal dissipation pathway during UTEM observations. The 1T-TaSe$_2$ samples with a thickness of about 50nm are suitable for the fs-laser excitation during ultrafast structural measurements with better stability. Firstly, the CDW modulations in these samples were checked using high-resolution STEM as typically shown in Fig.1b, in which the atomic structure associated with the commensurate modulations has been well identified in 1T-TaSe$_2$ as illustrated in the schematic structural models.

In previous literatures, measurements of transport and structural properties clearly revealed the presence of resistivity anomalies in correlation with the CDW transition in all TaSe$_2$ samples [13, 21]. For instance, Fig.1c shows the temperature dependence of resistivity for a TaSe$_2$ single crystal. Evident anomalous behavior in correlation with the CDW transitions is recognizable at about 470K. The ultrafast electron diffraction observations were performed under an acceleration voltage of 160 kV with a LaB$_6$ photocathode driven by 300-fs laser pulses (the wavelength: 347nm, the repetition rate: 100 kHz). The photo-induced phase transition was initiated by another fs pump laser (the wavelength: 520nm). Fig.2a shows three diffraction patterns obtained on a TaSe$_2$ single crystal at different time delays with pump fluence (F) of 2mJ/cm$^2$. Considering the weak reflecting intensity of the satellite spots, we firstly collected the ultrafast data with the electron bunches containing ~1000 electrons/pulse. These diffraction images are obtained respectively by integration of 10$^5$ electron pulses at the time 0, the positive 10ps, and the 20ps time delays. One of the most notable features as revealed in these diffraction patterns is the appearance of evident changes of CDW satellite spots in both their intensity and positions following with the photo excitation. As similarly discussed based on UED data [23-24], the temporal intensity decrease of the satellite spots under low laser fluences in this kind of layered CDW system can be well explained by the atomic motions resulting from the temporal evolution in the electronic spatial distribution and strong electron-phonon coupling. In present study, our UTEM investigations are focused mainly on the notable changes of CDW modulations associated with the phase transitions induced by laser excitation. So we display electron diffraction patterns in Fig. 2a taken from a TaSe$_2$ single crystal orientated along the [001] zone-axis direction, and they show up the main diffraction spots accompanying by six satellite spots that can be used to characterize temporal evolutions associated with the phase transition in TaSe$_2$. For facilitating the comparison, diffraction difference between the two positive frames of the 10ps and 20ps are also displayed in the right frame of Fig.2a, demonstrating a visible rotation for the CDW modulation. The rotation angle (~13$^o$) is fundamentally consistent with the data as thermally measured for the phase transition at T$_L$=470K.

For a better and clear view of these CDW dynamic features, we have theoretically analyzed all diffraction patterns using angular integration which yielding a one-dimensional curve in which the position shifts for CDW spots can be clearly displayed. Actually, it is also convenient to characterize the phase transition in TaSe$_2$ by using the modulation wave vector q (i.e. the order parameter) which changes accordantly as a function of the time-delay and excitation fluence. Alterations of wave vector and spot intensity can be directly measured from the relative change in comparison with the time = 0 frame as shown in Fig.2b-c. We firstly discuss the inter-conversion between two CDW states with different order parameters of $q_{IC}$ and $q_C$, it is notable in this CDW system that this non-equilibrium phase transition shows up an noteworthy "structurally isosbestic point" as clearly illustrated in Fig.2b, similar structural feature has been discussed in the previous literature for the La$_2$CuO$_{4+\delta}$ high-Tc superconductor in which the "isosbestic point" for the alteration of interlayer space is interpreted by photo-doping effects [25]. From structural point of view for this CDW phase transition, the existence of an isosbestic angle for the order parameter ($q_{iso}$) demonstrates that the $q_{IC}$ and $q_C$ phases yield their diffracting efficiencies in an equally ratio at this particular transient point [25-26]. Actually, we can described this phenomenon by a quite generally expression, the electron diffraction amplitude is the Fourier transform of lattice potential for the $q_{iso}$ reciprocal vector, $A(q_{iso})= \sum_{H,i} V_H \exp[2\pi i(H-q_{iso})\cdot r_i]$ (where the coefficients $V_H$ is lattice potential obtained from Fourier transform for the reciprocal lattice vector H). It is clearly recognizable that diffraction intensity is essentially governed by the alteration of the atomic position ($r_i$) in the transient state. Therefore, the presence of the "isosbestic point" demonstrate that the atomic structure in the crystal planes parallel to the vector $q_{iso}$ shows up very similar features in the IC-phase and C-phase. Based on our UTEM experimental measurements, we can estimate $q_{iso}$ using $q_{iso} \approx (2q_C+3q_{IC})/5$ as shown in Fig.2b，so the $q_{iso}$ vector could approximately go along the [1 2 0] zone axis direction and the crystallographically equivalent ones. This fact suggests that the (120) crystal planes could play a critical role for the understanding of the transient states in the CDW inter-conversion through a non-equilibrium phase transition. It is also proved that the coexistence and competition of IC-phase and C-phase can also be directly observed in the temporal evolution of microstructure, which depends evidently on the laser fluence and time delays.

We now go on to illustrate CDW dynamics accompanying the phase transition by using order parameters of $q_C$ and $q_{IC}$ for 1T-TaSe$_2$. So the decay of the initial $q_C$ structure and the formation of the $q_{IC}$ structure are quantified by integrating the intensity of relevant satellite spots, as illustrated by the intensity curves of Fig.2c. The red and blue curves are the total integrated intensity to the right and left of the crossing point in Fig. 2b, respectively, they directly give rise to the alternations of volume fraction of the initial $q_C$ phase and that of the $q_{IC}$ phase. In our measurements, the nature of the recovery process after this phase transition has been also examined, it is found that the restructuring process in many 1T-TaSe$_2$ crystals is relatively slow and often incomplete after 10 ns, so a longer time scale is required for properly stroboscopic observations. In our UTEM experiments, we have commonly performed the experimental measurements with the repeating frequency of less than 100 kHz and, frequently, an in-situ low temperature TEM holder is also used for getting better and stable results, it is shown that the 10μs between pulses is sufficient for the cooling of the 1T-TaSe$_2$ samples.

Considering the presence of visible space-charge effects in the ultra-fast imaging process, we have improved temporal resolution of experimental data by lowering the probe-laser power for photoemission. As a result, each electron pulse contains about 10 electrons, and the pulse duration is estimated to be about 1ps (FWHM) at the sample position as reported in our previous publication [12]. Fig.3a shows the experimental data with F = 1.8mJ/cm$^2$, each image for this ultrafast process is taken with an exposure time of 30 seconds. It is recognizable that satellite spots firstly decrease their intensity due to the CDW melting and charge reorganization, then the modulation wave moves towards a new direction following condensation of the $q_{IC}$ phase.

It is also known that the dynamical nature of the energy landscape in this CDW transition is determined by remarkable changes in the electronic distribution and by the electron-phonon interaction. Fig.3a depicts the observed trajectories in momentum-time space for two modulated structures with a clear q switch. In which the temporal evolution of the CDW satellite spots in the q-t space is shown as a three-dimensional plot (bottom). The initial structure of C-phase before time zero ($t_0$) is shown in the top right corner. After the phase transition, the new structure is indicated with a rotated CDW vector (top left corner). The change in these schematic pictures is exaggerated for illustrative purposes. In the q-t space picture, the intensity around the 30 degree represents the diffraction signal from C-phase, and the spot around 43 degree represents the IC-phase. We can also use a Gaussian function to fit the position and intensity of satellite spot at each time delay, as a result, the fraction of C-phase and IC-phase in the transient state can be obtained, as shown in Fig. 3b. It is notable that the CDW intensity in the C-phase decreases immediately following the ultrafast laser excitation. On the other hand, the intensity of IC-phase is almost unchanged until the time delay of 3ps. Our systematic analysis show that the major intensity conversion occur from 3ps to 6ps and slightly returned to an intermediate point from 6ps to 12ps, then it kept unchanged up to 1ns which is the large time delay in this measurement.

Fig. 3c depicts the change of peak intensity extracted from five sets experimental data. Three transient steps in the CDW transition are clearly recognizable as the reduction of C-phase, then the growth of IC-phase and then the partial recovery from IC-phase to C-phase, respectively. The recovery of the CDW amplitude occurs in transient process (step 3) can be well fitted using an exponential decay with a recovery time about 3ps. It is believed that the CDW recovery dynamics at this timescale dominated by the energy transport and conversion associated with acoustic phonon thermalization, this phenomenon is essentially in correlation with the anharmonic phonon decay which plays a critical role for understanding of the coupled electron–lattice order parameters as similarly discussed in ref. [23]. On the other hand, if the pump fluence is larger than $2mJ/cm^2$, our experiment results show that certain structural defects could appear in the CDW state, these defects can evidently influence the UTEM measurements due to their pinning effects on the inter-conversion of the CDW modulations.

Though measurements of physical properties at the CDW transitions in the TaSe$_2$

crystals have clearly demonstrated the first order characters in thermal equilibrium [13, 27-28], our UTEM study on the ultrafast structural changes of the order parameters often shows up rather different characters. Fig. 4a shows changes of order parameters as a function of laser fluence, illustrating a very complex nature for present non-equilibrium phase transition in 1T-TaSe$_2$. Actually, previous UED study showed that the laser excited 4H-TaSe$_2$ single crystals adopt a photo-induced phase transition with visible second-order characters. We firstly analysis the data obtained at the time delay of 20ps, as shown in the Fig.4a, indeed the results can be well fitted by I = tanh[1.78(Fc/F-1)$^{1/2}$], i.e. the well-known BCS approximation to the BCS gap equation for second-order phase transitions[29]. Moreover, we have also analyzed the transient features of CDW state at different time delays, it is noted that the spot intensity changes with laser fluence often appear at regions between the first and second order transition, as clearly illustrated by data obtained at T = -20ps (blue curves), this feature is believed to be arising partially from lattice thermalization and structural alteration with a tendency of the first order transitions. It is also noted that in general the intensity of CDW spots often changes nonlinearly with the laser fluence (F), we can see a threshold fluence at about ~ 0.75mJ/cm$^2$, and diffraction intensity above this threshold has an apparent nonlinear decrease with the laser fluence rise. Moreover, as the fluence is larger than 2.2mJ/cm$^2$, the defect structures and twining can commonly appears in the samples, which could yield irreversible features in the UTEM experiments.

In summary, the present UTEM work illustrates the importance of directly observing fast changes of the CDW modulation and related atomic motions for the non-equilibrium phase transition, which provide a rich variety of microstructural information for understanding the transient states on strongly correlated lattice dynamics. Through the photoexcitation, the C-phase transform to the IC-phase is observed in the time-scale of about 3 ps, and remarkable fast changes on the directions of CDW waves are clearly identified correlation with CDW inter-conversion and charge re-condensation. Moreover, the transition of superstructure shows up a notable "structurally isosbestic point" resulting from the competition/coexistence of two CDW phases during this non-equilibrium phase transition. Certain fundamental features and second order characters in the transient states have been also analyzed for the photo-induced CDW transitions. Our study demonstrated that, with further UTEM improvements of spatial/temporal resolutions and increase in the signal-to-noise ratio, this advance technique could be used to find signatures which are critical for address the specific charge and atomic motions during the CDW phase transitions.

## Acknowledgements

This work was supported by National Basic Research Program of China 973 Program (Grant Nos. 2011CBA00101, 2010CB923002, 2011CB921703, 2012CB821404, 2011CBA00111), the Natural Science Foundation of China (Grant Nos. 11274368, 51272277, 11074292, 11004229, 11190022, U1232139), and Chinese Academy of Sciences.

————————

[*]These authors contributed equally to this work.
Corresponding author:
[†]hxyang@aphy.iphy.ac.cn
[‡]ljq@aphy.iphy.ac.cn

**Figure captions**

FIG. 1 Photograph of the 4D UTEM at IOP，CAS, together with the relevant experimental data. (a) A photograph of the UTEM at IOP, the inserted image is the 1T-TaSe$_2$ single crystals mounted on well-aligned MWCNT bunches. (b) High-resolution STEM image for the CDW modulation in 1T-TaSe$_2$, the inserted image is schematic structural model for commensurate phase. (c) Temperature dependence of resistivity for the 1T-TaSe$_2$ shows clear phase transition at 470K. The inserted images are electron diffraction of C-phase and IC-phase, respectively.

FIG. 2 Time-resolved experimental data for CDW phase transition of 1T-TaSe$_2$ with a fluence of 2.0 mJ/cm$^2$, showing the "structurally isosbestic point". (a) Temporal frames of diffraction patterns for time delays of 0ps, 10ps and 20ps, respectively. Diffraction difference clearly illustrates the rotation of the satellite reflections. (b) Angular integrated 1D diffraction curves for different time delays. (c) The diffraction intensity change of C-phase and IC-phase as a function of time delay.

FIG. 3 Dynamic features and temporal evolution of the C-phase state to IC-phase state in 1T-TaSe$_2$ on photoexcitation. (a) The alternation CDW with the pump fluence of 1.8mJ/cm$^2$. (b) The position and intensity changes of satellite spots as a function of time delay, as obtained by fitting with the Gaussian function. (c) Transient states reveal a 3-step process of phase transition, i.e. the reduction of C-phase, the increase of IC-phase, and partial recovery from IC-phase to C-phase.

FIG. 4 Fluence dependence of transient diffraction signal. (a) Changes of order parameters as a function of laser fluence at time delay of t = 20ps (positive) and -20ps (negative). The structural evolution at the time delay of 20ps can be well fitted by a BCS function, showing typical characters of the second-order phase transition. However, the structural evolution at the time delay of -20ps appears at the region

between the first and second order transition. (b) The changes of ultrafast diffraction difference as a function of laser fluence. It shows a threshold, a nonlinearly increase region and an irreversible region.

**FIG. 1**

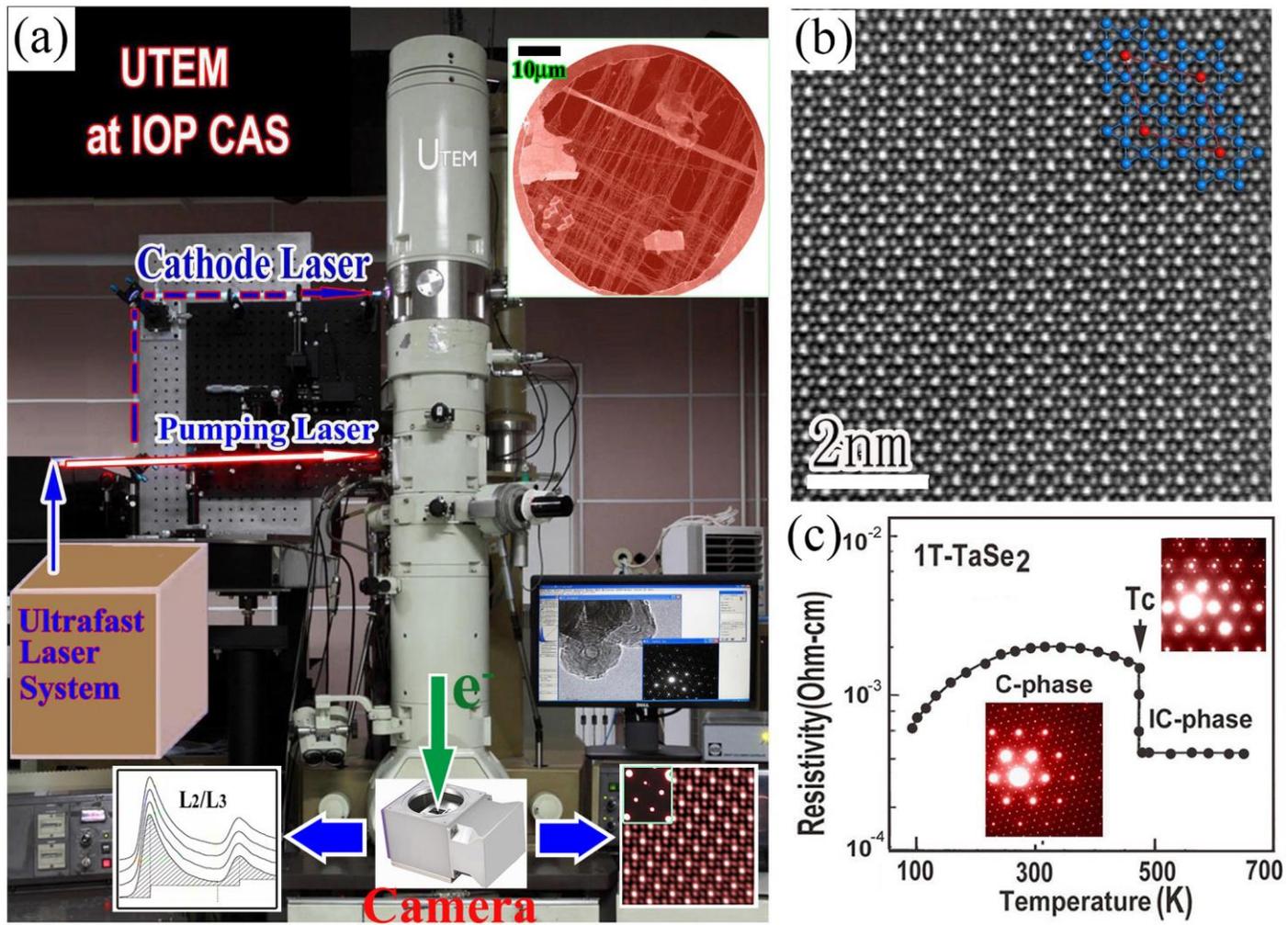

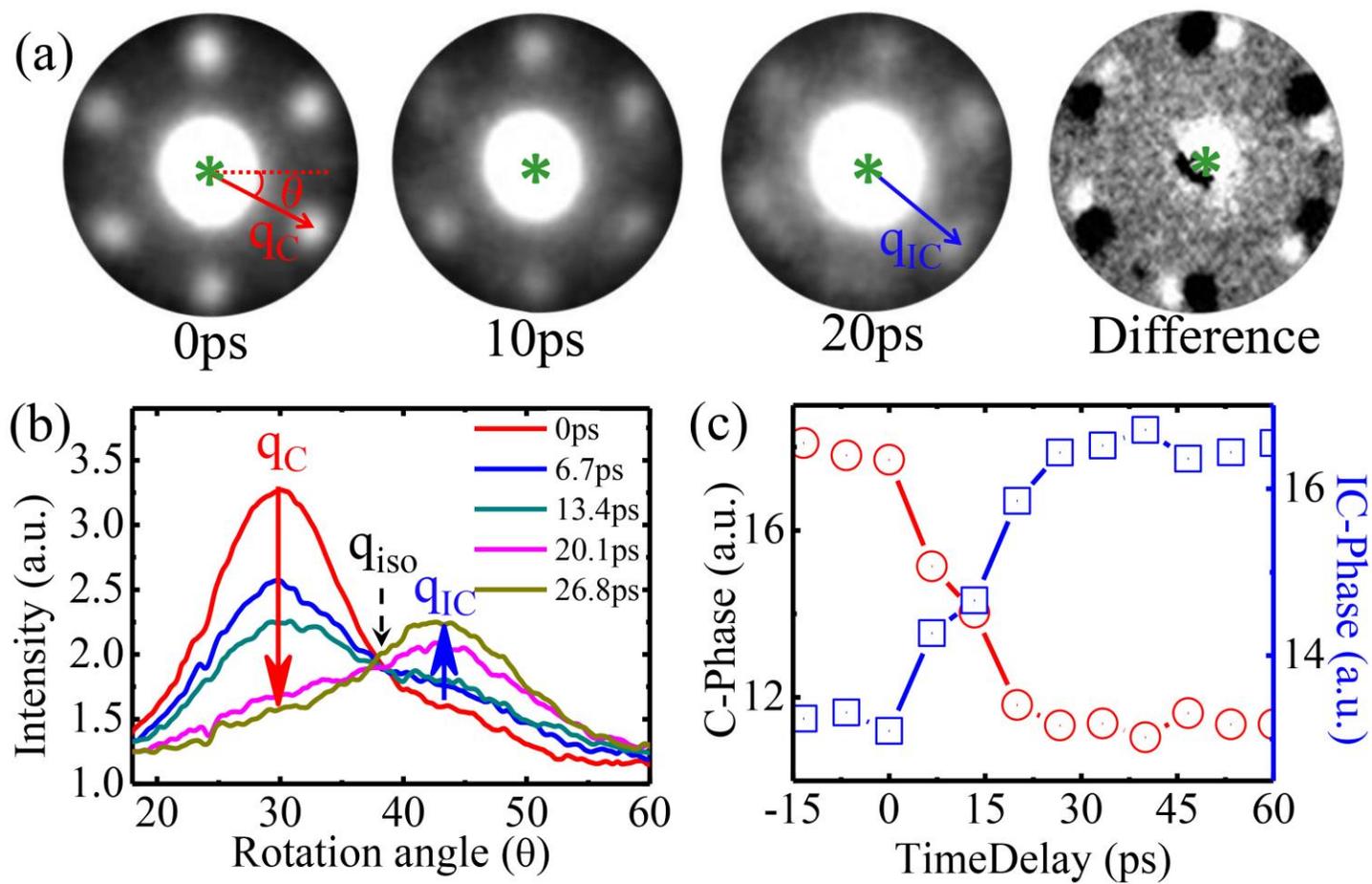

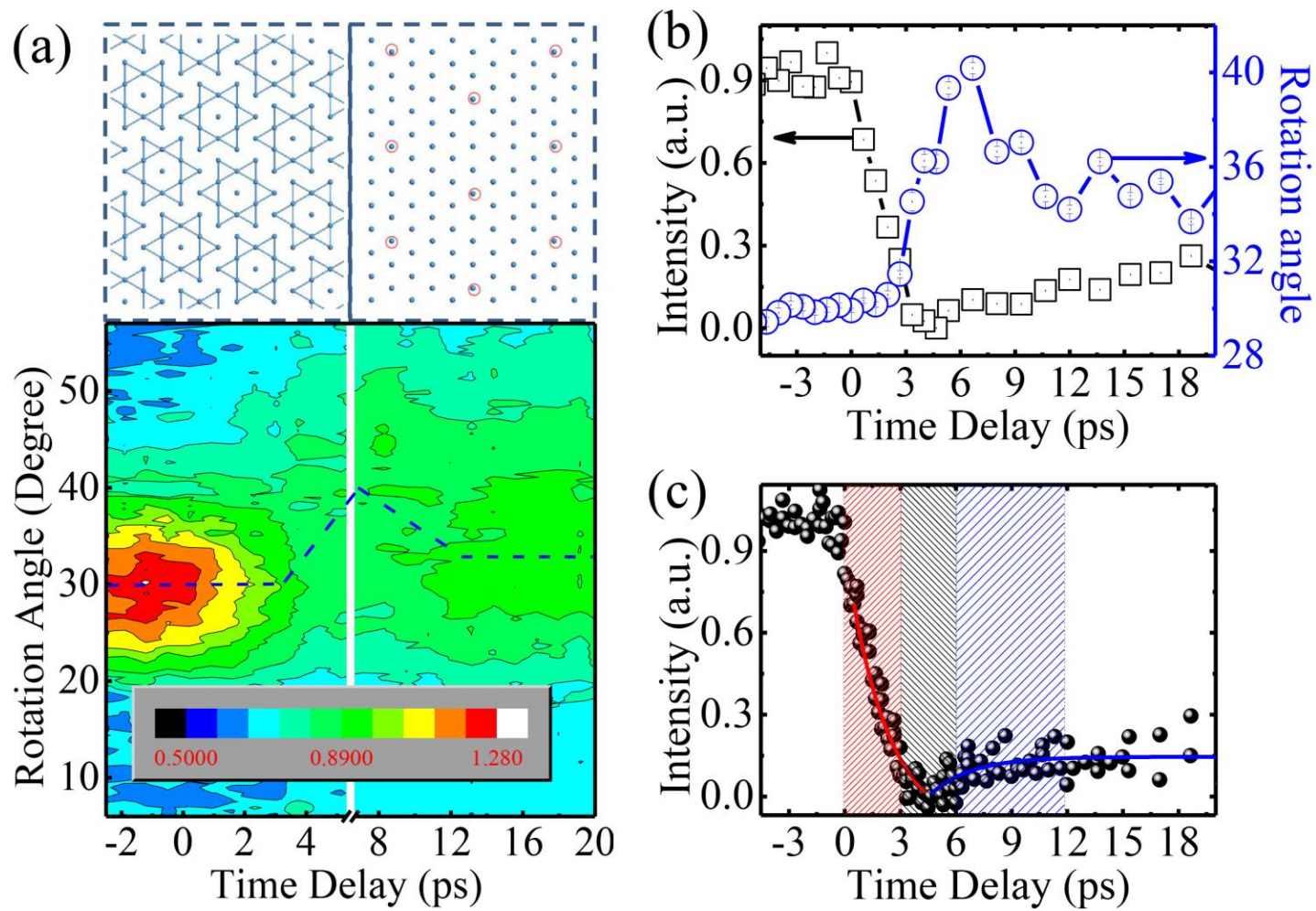

**FIG. 4**

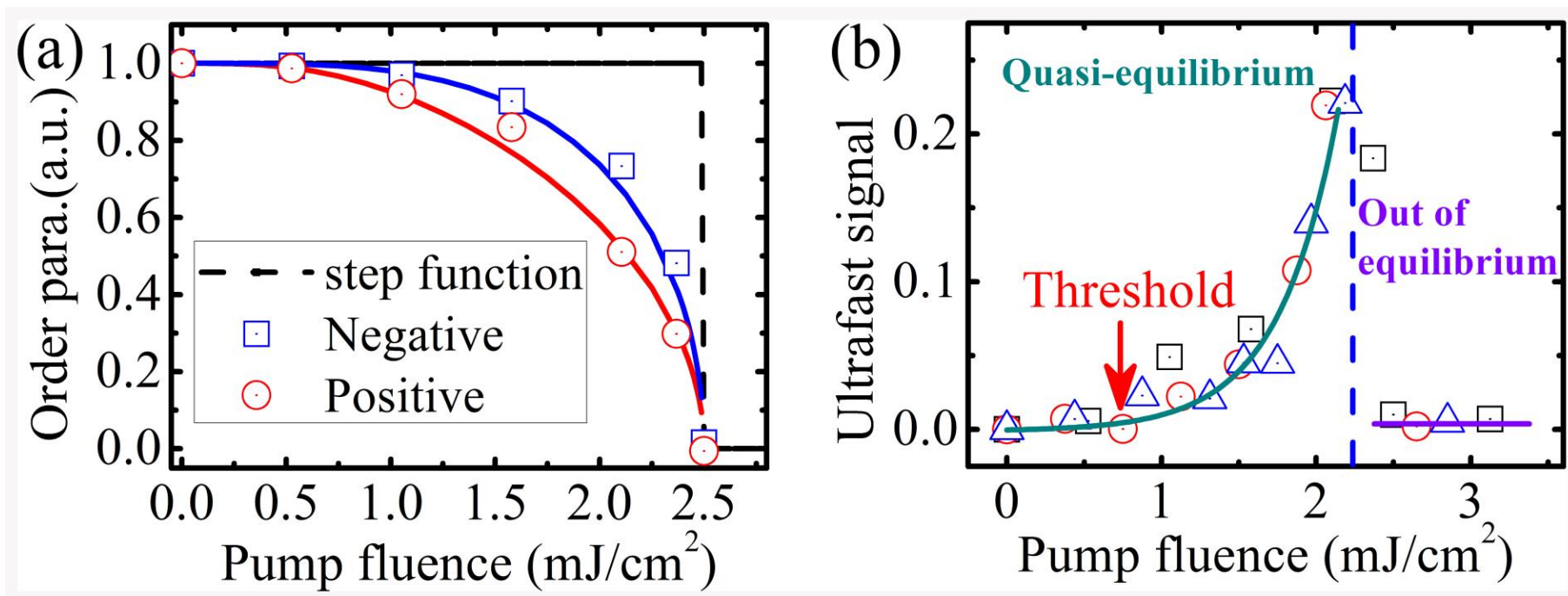